\documentclass[runningheads]{llncs}
\usepackage{mathpartir}  
\usepackage{xspace,graphicx,prettyref,subfiles}
\usepackage{xcolor}
\usepackage{amssymb}   
\usepackage{url}
\usepackage{wrapfig}
\makeatletter
\AtBeginDocument{%
  \def\doi#1{\url{https://doi.org/#1}}}
\makeatother

\newrefformat{eq}{Eq.\ \textup{(\ref{#1})}}
\newrefformat{lem}{Lem.\ \ref{#1}}
\newrefformat{def}{Def.\ \ref{#1}}
\newrefformat{thm}{Thm.\ \ref{#1}}
\newrefformat{cha}{Ch.\ \ref{#1}}
\newrefformat{sec}{Sec.\ \ref{#1}}
\newrefformat{tab}{Tab.\ \ref{#1}}
\newrefformat{fig}{Fig.\ \ref{#1}} 

\usepackage{amsmath}  
\usepackage{amsfonts}  
\usepackage{stmaryrd}  

\newcommand{\rref}[1]{\prettyref{#1}}
\newcommand{\produces}{{:}{:}{=}}
\newcommand{\alternative}{~|~}

\newcommand{\D}[1]{{#1}'}
\newcommand{\der}[1]{({#1})'}

\newcommand{\KeYmaeraX}{KeYmaera\ X\xspace}
\newcommand{\dL}{\(\mathsf{dL}\)\xspace}

\newcommand{\ptest}[1]{?{#1}}

\newcommand{\prepeat}[1]{{#1}^*}
\newcommand{\pevolve}[1]{\{{#1}\}}
\newcommand{\pevolvein}[2]{\{{#1} \& {#2}\}}

\newcommand{\passign}[2]{#1\mathrel{{:}{=}}#2}

\newcommand{\dbox}[2]{[{#1}]{#2}}
\newcommand{\ddiamond}[2]{\langle{#1}\rangle{#2}}
\newcommand{\limply}{\rightarrow}

\newcommand{\kwif}{\mathsf{if}}
\newcommand{\kwelse}{\mathsf{else}}
\newcommand{\kwwhile}{\mathsf{while}}

\newcommand{\vamts}{\mathsf{amts}\xspace}
\newcommand{\vka}{\mathsf{k}_A\xspace}
\newcommand{\vkb}{\mathsf{k}_B\xspace}
\newcommand{\vkc}{\mathsf{k}_C\xspace}
\newcommand{\vkf}{\mathsf{k_F}\xspace}
\newcommand{\vkr}{\mathsf{k_R}\xspace}
\newcommand{\vkt}{\mathsf{k_T}\xspace}
\newcommand{\vkra}{\mathsf{k_{ra}}\xspace}
\newcommand{\vkrb}{\mathsf{k_{rb}}\xspace}
\newcommand{\va}{\mathsf{A}\xspace}
\newcommand{\vb}{\mathsf{B}\xspace}
\newcommand{\vc}{\mathsf{C}\xspace}
\newcommand{\ve}{\mathsf{E}\xspace}
\newcommand{\veu}{\mathsf{U}\xspace}
\newcommand{\vek}{\mathsf{KE}\xspace}
\newcommand{\veps}{\epsilon\xspace}
\newcommand{\vrt}{\mathsf{T}\xspace}
\newcommand{\vhpt}{\mathsf{T}_s\xspace}
\newcommand{\vtmax}{\mathsf{T}_{\mathit{max}}\xspace}
\newcommand{\vequiA}{\tilde{A}\xspace}
\newcommand{\vequiB}{\tilde{B}\xspace}
\newcommand{\vainit}{\mathsf{A}_{0}\xspace}
\newcommand{\vbinit}{\mathsf{B}_{0}\xspace}

\newcommand{\vamid}{\mathsf{A}_{1}\xspace}
\newcommand{\vbmid}{\mathsf{B}_{1}\xspace}

\newcommand{\veinit}{\mathsf{E}_0\xspace}
\newcommand{\vison}{\ensuremath{\mathsf{isOn}}\xspace}
\newcommand{\vconst}{\mathsf{const}\xspace}

\newcommand{\vode}{\mathsf{ode}\xspace}

\newcommand{\vctrl}{\ensuremath{\mathsf{ctrl}}\xspace}
\newcommand{\vtaylorHi}{\mathsf{taylor}^+\xspace}
\newcommand{\vrate}{\mathsf{rate}\xspace}

\begin{document}
\title{Chemical Case Studies in KeYmaera X}
\author{Rose Bohrer\orcidID{0000-0001-5201-9895}} 
\authorrunning{R.\ Bohrer}
\institute{Worcester Polytechnic Institute, Worcester MA 01609, USA\\
\email{rose.bohrer.cs@gmail.com}}
\maketitle              
\begin{abstract}
Safety-critical chemical processes are the backbone of multi-billion-dollar industries, thus society deserves the strongest possible guarantees that they are safe.
To that end, models of chemical processes are well-studied in the formal methods literature, including hybrid systems models which combine discrete and continuous dynamics.
This paper is the first to use the \KeYmaeraX theorem-prover to verify chemical models with differential dynamic logic.
Our case studies are novel in combining the following: we provide strong general-case correctness theorems, use particularly rich hybrid dynamics, and have particularly rigorous proofs.
This novel combination is made possible by \KeYmaeraX.

Simultaneously, we tell a general story about \KeYmaeraX: recent advances in automated reasoning about safety and liveness for differential equations have enabled elegant proofs about reaction dynamics.
\end{abstract}

\section{Introduction}
\label{sec:introduction}
Modern industry relies critically on all kinds of chemical processes: some occur in computer-controlled reactors, some occur free of control.
Chemical engineering has provided many classical insights about both: safe and optimal control~\cite{glad2018control} of reactors~\cite{cpc} is a field in its own right, as are reaction kinematics (dynamics) even in the absence of control~\cite{hellgardt2000engineering}.

Because both controlled and uncontrolled reactions are crucial, we consider both: an irreversible exothermic reaction with a model-predictive bang-bang controller (\rref{sec:irreversible}) and an uncontrolled reversible reaction (\rref{sec:reversible}).
Both have verification challenges which make for good benchmark problems.
The nonreversible reaction's nuanced dynamics entail nontrivial correctness arguments for model-predictive controllers.
The reversible reaction's long-term asymptotic behavior, though classic, tests the ability of current-generation tools to verify asymptotic properties, e.g., stability~\cite{DBLP:conf/cdc/MitraL04} or persistence~\cite{sogokon2017verifying}.

Safe reactions are crucial to human safety. Properties like persistence, stability, and optimality are crucial to human productivity. Thus, formal methods for chemical reactions are extensively studied~\cite{bauer2000case,ctx4405807090004746,lukoschus2004compositional,hassapis1998validation,DBLP:series/asc/NellenAW15}.

To our knowledge, however, the reaction models and proofs presented here are the first-ever in a \emph{hybrid systems theorem prover}.
Specifically, we use the \KeYmaeraX~\cite{DBLP:conf/cade/FultonMQVP15} prover for \emph{differential dynamic logic} (\dL)~\cite{Platzer18} to achieve a unique combination of expressive dynamics, general-case guarantees, and rigor for the first time.
The tradeoffs between theorem-proving and other formal methods are well-known; see \rref{sec:related-work} for detailed discussion.

Our contribution was enabled by new stability~\cite{DBLP:journals/fac/TanP21}, 
variant~\cite{DBLP:journals/fac/TanP21}, and Darboux polynomial~\cite{DBLP:journals/jacm/PlatzerT20} proof tools in \KeYmaeraX, simplifying our proof arguments.
Our case studies make essential use of these features and thus demonstrate the impacts of the latest advances in proof automation.

\section{Background}
\label{sec:background}
All our proofs are computer-checked in the \KeYmaeraX prover, which carefully prevents the use of unsound reasoning~\cite{DBLP:conf/cpp/BohrerRVVP17}.
This rigor is crucial in practice: many techniques used here had predecessors~\cite[Table\ 1]{DBLP:journals/fac/TanP21} which were found to be unsound, which is unacceptable for safety-critical systems.

In \KeYmaeraX, correctness properties are stated and proved in \emph{differential dynamic logic} (\dL)~\cite{Platzer18}, where hybrid systems are written in \emph{hybrid program} notation.
We discuss \dL, then \KeYmaeraX usage.

\subsection{Differential Dynamic Logic}
\label{sec:background-ddl}
We provide a primer on \dL syntax and semantics (meaning); see the literature~\cite{Platzer18} for details.
Semantics are state-based: a state $\omega$ maps every variable $x$ to a real-number value $\omega(x) : \mathbb{R}$.
The syntax consists of terms (with a numeric meaning in each state), hybrid programs (which can nondeterministically change the state when run), and formulas (which are true or false in each state).
Hybrid programs and formulas may both contain each other.
We use standard notation to define syntax, e.g., $B \mathrel{\produces}  C \alternative  D$ means every $B$ is either a $C$ or a $D$.

\begin{definition}[Terms]
Terms $e, \tilde{e}$ of \dL are defined by:%
\[e, \tilde{e} \mathrel{\produces} q \alternative x \alternative e + \tilde{e} \alternative e \cdot \tilde{e}\qquad\qquad\text{where $q \in \mathbb{Q}$}\]%
\label{def:dl-terms}%
\end{definition}%
\vskip-0.3in Rational-valued literal numbers are written $q$. 
Real-valued variables are written $x$.
Sum  $e + \tilde{e}$ is the sum of terms $e$ and $\tilde{e}$. 
Product $e \cdot \tilde{e}$ is the product of $e$ and $\tilde{e}$. 
In every state, the meaning of every term is some real number.

\begin{definition}[Hybrid Programs]
Hybrid programs $\alpha, \beta$ are defined by: 
\[\alpha, \beta \mathrel{\produces} ~\ptest{P} \alternative \passign{x}{e} \alternative \pevolvein{\D{x}=f(x)}{Q}
\alternative \alpha \cup \beta \alternative \alpha;\beta \alternative \prepeat{\alpha}\]
\label{def:dl-hybrid-programs}
\end{definition}
\vskip-0.3in Hybrid programs are defined by their \emph{runs}: from a starting state, what final states are reachable?
Hybrid programs can have one run (deterministic), many runs (nondeterministic), or zero runs (early termination).
Programs $\ptest{P}$ and $\pevolvein{\D{x}=f}{Q}$ contain formulas $P$ and $Q$; see \rref{def:dl-formulas} for more about formulas.

The test program $\ptest{P}$ never modifies the state; if formula $P$ is true, then $\ptest{P}$ ends in the current state, but if $P$ is false, then $\ptest{P}$ has no final states, representing execution failure.
Deterministic assignment $\passign{x}{e}$ updates the state by storing the current value of term $e$ in variable $x$.
Ordinary differential equation systems (ODEs) are the defining feature of hybrid programs: ODEs composed with discrete operations model hybrid systems.
ODE $\pevolvein{\D{x}=f(x)}{Q}$ evolves in continuous time with $\D{x}=f(x),$ where $f(x)$ is a term.
The duration of evolution is nondeterministic.
If an \emph{evolution domain constraint} $Q$ is provided, $Q$ is tested continuously, and evolution must stop before $Q$ ever becomes false.
Choices $\alpha \cup \beta$ nondeterministically run \emph{either} $\alpha$ \emph{or} $\beta,$ as opposed to running both.
Composition $\alpha;\beta$ runs $\alpha,$ then $\beta$ in the resulting state(s).
Duration of loops $\prepeat{\alpha}$ is nondeterministically-chosen but finite: zero, one, or many repetitions can occur.
If desired, standard conditional and looping constructs are derivable (where $P$ is a formula, $\lnot P$ is its negation, and $\alpha$ is a hybrid program):
\begin{align*}
\kwif(P)\{\alpha\}\kwelse\{\beta\} &\equiv \{\ptest{P};\alpha\}\cup\{\ptest{\lnot P};\beta\}\\
\kwwhile(P)\{\alpha\} &\equiv \prepeat{\{\ptest{P};\alpha\}};\ptest{\lnot P}
\end{align*}

\begin{definition}[Formulas]
There are many formulas $P, Q$ in \dL. We only use:

\[P, Q \mathrel{\produces} 
\cdots \alternative e \geq \tilde{e} \alternative  \lnot P \alternative P \land Q \alternative P \limply Q \alternative \dbox{\alpha}{P} \alternative \ddiamond{\alpha}{P}
\]
\label{def:dl-formulas}
\end{definition}
\vskip-0.3in Formulas represent true/false questions about the state $\omega$.
Comparison $e \geq \tilde{e}$ is true whenever the value  of $e$ is at least that of $\tilde{e}$ in a given state.
All other comparisons $e>\tilde{e}, e=\tilde{e}, e\neq\tilde{e},  e\leq\tilde{e}, e<\tilde{e}$ are definable using $e\geq\tilde{e}$ and other logical connectives, so we use them freely.
Negation $\lnot P$ is true when $P$ is false.
Conjunction $P \land Q$ is true when both $P$ and $Q$ are.
Implication $P \limply Q$ is true when $P$'s truth would imply $Q$'s truth.

The defining formulas of \dL, $\dbox{\alpha}{P}$ and $\ddiamond{\alpha}{P},$
are respectively true in state $\omega$ if \emph{every} or \emph{some} of $\alpha$ starting from state $\omega$ ends in a state where $P$ is true. 

When $\alpha$ is an ODE, \emph{all runs} equates to \emph{all time}, e.g., these readings apply:
\begin{itemize}
\item $P \limply \dbox{\alpha}{Q}$ assumes $P$ at first, then proves $Q$ forever
\item $P \limply \ddiamond{\alpha}{Q},$ assumes $P$ at first, then proves $Q$ eventually 
\item $P \limply \ddiamond{\alpha}{\dbox{\alpha}{Q}},$ assumes $P$ 
at first, then proves $Q$ eventually becomes true, then stays true forever.
\end{itemize}

\KeYmaeraX proves truth in \emph{every state}, called \emph{validity}.
\begin{definition}[Validity]
A \dL formula is \emph{valid} if it is true in every state.
\label{def:validity}
\end{definition}

We use standard notation for axioms and proof rules.
\begin{definition}[Proof Rules]
Each rule has a horizontal line and means: if all \emph{premise} formulas above the line are valid, so is the \emph{conclusion} formula below the line.
Rules can use \emph{schema variables} such as $P$ or $\alpha$ when the rule applies to all programs or formulas, respectively.
\label{def:axioms-rules}
\end{definition}
For example, the \textsf{loop} rule
\[\inferrule*[Left=\textsf{loop}]{P \limply J \\ J \limply \dbox{\alpha}{J} \\ J \limply Q}{P \limply \dbox{\prepeat{\alpha}}{Q}}\]
means for all $P, Q, J, \alpha$ that if premises $P \limply J,$ $J \limply \dbox{\alpha}{J},$ and $J \limply Q$ are all valid, so is $P \limply \dbox{\prepeat{\alpha}}{Q}$.
Formula $J$ is \emph{proved} true for all iterations, thus we call $J$ the \emph{loop invariant}.
This \emph{proven} loop invariant should not be confused with use of the word \emph{invariant} in hybrid automata  to mean an \emph{assumed} constraint on ODE evolution. We call such constraints \emph{evolution domain constraints}.


\subsection{\KeYmaeraX}
\label{sec:background-kyx}
\begin{wrapfigure}{r}{0.5\textwidth} 
    \centering
    \includegraphics[width=0.5\textwidth]{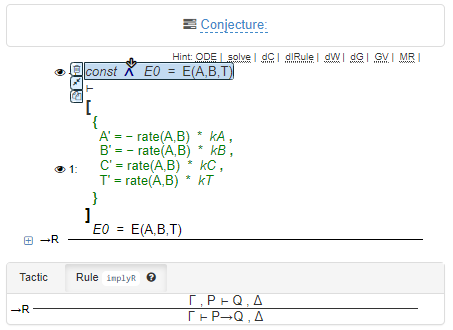}
    \caption{\KeYmaeraX screenshot. Clicking the highlighted symbol performs a proof step. The last proof rule is shown at bottom. Recommended proof steps are displayed as hints.}
    \label{fig:kyx-screenshot}
\end{wrapfigure}

We briefly discuss the user experience of \KeYmaeraX~\cite{DBLP:conf/fide/MitschP16}.
The user interface is displayed in \rref{fig:kyx-screenshot}.
\KeYmaeraX is an interactive, tactic-based prover.
This means that the user interactively tells the prover which proof technique to use, but each technique is implemented as a \emph{tactic}~\cite{DBLP:conf/itp/FultonMBP17}, i.e., a program.
A proof technique can be a simple, specific rule or a complex proof search procedure. 
For example, there is a \emph{default} (or \texttt{auto}) proof procedure which attempts many proof techniques and can solve many simpler problems fully-automatically.
In summary, the amount of user effort can vary greatly between proofs. 
Throughout this paper, we will discuss the level of interaction needed for each proof and discuss how new rules and automation helped keep the level of user effort manageable.
The tactic-based approach also means that no matter how complex proof methods are, they are implemented using simple steps from the small trusted core of the prover, thus proofs stay rigorous.

\section{Results}
\label{sec:results}
We contribute case studies on two classic kinds of chemical reactions.
The first is an irreversible reaction in a well-mixed adiabatic batch reactor, which we chose because batch reactors~\cite[\S2.10]{hellgardt2000engineering} are a foundational technology for chemical plants throughout industry.
The second case study is a reversible reaction between two compounds, i.e., where the output can react again and form the input.
We chose reversible reactions because they too are essential to industry.
Notably, ammonia synthesis is a reversible reaction that provides the backbone for modern fertilizer-based, industrial-scale agriculture~\cite{liu2014ammonia}.
Both case studies emphasize recent advances in \KeYmaeraX proof automation, which contributed to highly general results.
Remaining limits on generality are discussed in each subsection.


\subsection{Controlled Irreversible Reactions}
\label{sec:irreversible}
We formalize a classic scenario: an irreversible, exothermic reaction in an adiabatic, well-mixed batch reactor.
\emph{Irreversible}~\cite[\S2.1]{hellgardt2000engineering} means the reaction is one-way: outputs do not react to create inputs. 
\emph{Adiabatic}~\cite[\S2.14]{hellgardt2000engineering} means heat does not leave or enter the reactor.
\emph{Well-mixed}~\cite[\S2.12]{hellgardt2000engineering} means the reaction occurs evenly in space throughout the reactor.
In this basic synthesis reaction, two (first-order) reactants react to form a third, plus heat:
\[A + B \longrightarrow C + \mathit{heat}\]

The case study contains four models, each with proof.
The first shows conservation of energy, validating that adiabatic reactors are closed systems.
The remaining three models add a model-predictive bang-bang controller~\cite{glad2018control}, which predicts future behavior according to the model, then applies an all-or-nothing control action.
It is proved that the control ensures a safety property: overheating is prevented.
We use this standard control approach in order to focus on the continuous reaction dynamics.
The driving difference between the last three models is their increasingly complex reaction dynamics, which mandate increasingly complex controls and proofs.
In the second model, the reaction rate is constant.
In the third model, the rate depends linearly on temperature, changing exponentially with respect to time.
In the final model, the rate is proportional to the product of temperature and each concentration, with resulting dynamics beyond a simple exponential, yet still approximate.
Approximate results are the best that can be expected. 
We discuss why, including verification challenges.

Each model approximates textbook~\cite[Eq.\ 2.93]{hellgardt2000engineering} reaction dynamics, where the reaction $\mathit{rate}$ is proportional to the product of concentrations of each reactant $A$ and $B$ multiplied by a coefficient. Recall that the \emph{concentration} of a reactant in a mixture is the quantity of that reactant per unit quantity of the mixture.
The rate equation is  $\mathit{rate} = k A B$
where $k$ is an exponential given by the Arrhenius equation~\cite[Eq.\ 5.1]{hellgardt2000engineering}.
That is, $k(T) = k_0 e^{-E/RT}$ where $T$ is temperature, $R$ is the ideal gas constant, $E$ is the reaction's activation energy and $k_0$ a constant.

Analysis of the reaction rate dynamics is nontrivial: $\mathit{rate}$ is a product of three continuously-changing quantities, resulting in a non-linear ODE.
Moreover, $k(T)$ is exponential in $T,$ resulting in a \emph{non-polynomial} ODE.
\KeYmaeraX handles non-linear ODEs well, but is restricted to polynomial ODEs, as is standard.
We thus reach out first limitation: to ensure a polynomial ODE, we approximate the temperature dependence as linear.
This assumption is reasonable because polynomial ODEs are a standard assumption, and our nonlinear dynamics are still richer than prior models
~\cite{DBLP:conf/hybrid/QiwenW95,ctx4405807090004746,lukoschus2004compositional,hassapis1998validation,DBLP:series/asc/NellenAW15}.
Our second limitation is that the reactants are first-order, so their influence on $\text{rate}$ is linear.
We do so because such reactions are common and lead to elegant equations.
\KeYmaeraX supports polynomials of any degree, so we expect the approach to work for higher-order reactions, so long as the order is fixed.
Notwithstanding these limitations, the results are fully general in the sense that they are fully parametric, e.g., the results can be applied to \emph{any} reactants in \emph{any} amount by plugging in new coefficients and concentrations.

\subsubsection{Energy Conservation}
The basic \dL model for energy conservation is presented in \rref{fig:irreversible-conserve}.
Energy conservation is interesting in its own right, because it implies the system is closed.
This helps support our claim that the model is \emph{adiabatic}: heat energy does not leave nor enter.
\begin{figure}[!hbtp]\vskip-0.3in
\begin{align*}
&\ve \equiv \vek + \veu\quad \veu \equiv \min(\va/\vka,\vb/\vkb)\,\vkt\quad\vek \equiv \vrt\quad\vrate \equiv \vhpt\,\vainit\,\vbinit\,\vkra + \vkrb\\
&\vconst \equiv \vkra > 0 \land \vkrb \geq 0 \land \vka > 0 \land \vkb > 0 \land \vkc > 0 \land \vkt > 0\\
&\vode \equiv \pevolve{\D{\va} = -\vrate\,\vka, \D{\vb} = -\vrate\,\vkb, \D{\vc} = \vrate\,\vkc, \D{\vrt} = \vrate\,\vkt}\\[0.1in]
&(P \limply \dbox{\alpha}{Q}) \equiv (\vconst \land \veinit = \ve \limply \dbox{\vode}{\veinit = \ve})
\end{align*}
\caption{Basic irreversible model conserves energy}
\label{fig:irreversible-conserve}
\end{figure}\vskip-0.1in
The variables $\va, \vb,$ and $\vc$ stand for the current concentration of each reactant present in the reactor.
Reactor temperature is written $\vrt$.
In our analysis, we decompose energy into kinetic (heat) and potential (chemical) energy: 
$\ve \equiv \vek + \veu.$
Potential energy $\veu \equiv \min(\va/\vka,\vb/\vkb)\,\vkt$ is the product of the amount (concentration) of $\vc$ remaining to be produced (the reaction ends when either $\va$ or $\vb$ is exhausted) with the heat released per unit amount ($\vc$).
That is, we model $\vc$ as if it possesses no potential energy, since we are interested only in energies relevant to the current reaction.
We model the reaction rate as  $\vhpt\,\vainit\,\vbinit\,\vkra + \vkrb,$ which makes two intentional simplifications.
First, we use approximate \emph{current} concentrations $\va, \vb$ with \emph{initial} concentrations $\vainit, \vbinit.$
Secondly, we simplify the temperature factor to $\vhpt,$ which is a \emph{constant} even as temperature $\vrt$ changes, thus the influence of heat is \emph{static} throughout the reaction.
We determine the reaction rate as a product of the concentration factor and temperature factor.
For generality, the coefficients $\vkra, \vkrb$ let the rate be any \emph{linear function of} the product.
Formula $\vconst$ simply specifies the signs of constants.

The $\vode$ indicates that all concentrations $\va, \vb, \vc$ and the reactor temperature $\vrt$ all change proportional to the reaction rate; $\va$ and $\vb$ are lost as $\vc$ and heat are gained.
Coefficients $\vka,\vkb,\vkc,\vkt$ indicate the rates at which each changes, which may depend respectively on the stoichiometric coefficients of the reaction or how strongly exothermic it is.

Finally, the theorem statement $(P \limply \dbox{\alpha}{Q})$ states that under the simple $\vconst$ant assumptions, energy is conserved because at all times the current energy $\ve$ remains equal to its initial value $\veinit$.
We now describe the proof of the theorem in \KeYmaeraX.

\paragraph{Proof.}
The default proof procedure of \KeYmaeraX (\rref{sec:background-kyx}) proves the theorem automatically with \emph{differential invariants}~\cite[Lem. 11.3]{Platzer18}, demonstrating the capabilities of this standard \dL rule.
We present (the relevant case of) \emph{differential invariant}~\cite[Lem. 11.3]{Platzer18} rule
\[\inferrule*[Left=DI]
{Q \limply \dbox{\passign{\D{x}}{f(x)}}{\der{e} = \der{\tilde{e}}}}
{e=\tilde{e}\limply \dbox{\pevolvein{\D{x}=f(x)}{Q}}{e=\tilde{e}}}\]
which shows $e=\tilde{e}$ is true throughout an ODE if it holds initially and differentials are equal throughout.
We prove $\veinit = \ve$ thus: $\veinit$ is constant, so proving $\D{\ve}=0$ throughout suffices.
 Expanding the definition of $\ve$ yields $\der{\ve} = \der{\vrt + \min(\va/\vka,\vb/\vkb)\,\vkt}
= \vrate\,\vkt + \min(\der{\va}/\vka,\der{\vb}/\vkb)\,\vkt
= \vrate\,\vkt + \min(-\vrate\,\vka/\vka,-\vrate\,\vkb/\vkb)\,\vkt
= \vrate\,\vkt + \min(-\vrate,-\vrate)\,\vkt
= (\vrate - \vrate)\,\vkt = 0$.
Due to \KeYmaeraX's automation, the entire proof is automatic.

\newcommand{\grayout}[1]{{\color{black!70}{#1}}}

\subsubsection{On-Off Reactions}
This model keeps the basic heating dynamics but adds bang-bang control.
\rref{fig:irreversible-bangbang} describes the model in full.
Parts unchanged from \rref{fig:irreversible-conserve} are \grayout{grayed out} to aid comparison.
The impact of this theorem is that the reactor is provably safe under idealistic assumptions, i.e., when concentrations and temperatures change very little or have little impact on reaction rate.
\begin{figure}[!hbtp]\vskip-0.3in
\begin{align*}
&\grayout{\vrate \equiv \vhpt\,\vainit\,\vbinit\,\vkra + \vkrb}\\
&\vconst \equiv \grayout{\vkra > 0 \land \vkrb \geq 0 \land \vka > 0 \land \vkb > 0 \land \vkc > 0 \land \vkt > 0} \land \vrt >0 \land \veps > 0\\
&\vctrl \equiv \{\kwif(\vtmax - \vrt \leq \veps\,\vrate\,\vkr)\{\passign{\vison}{0}\}\kwelse\{\passign{\vison}{1}\}\};\passign{t}{0}\\
&\vode \equiv \{\D{\va} = \vison \cdot \grayout{-\vrate\,\vka}, \D{\vb} = \vison \cdot\grayout{-\vrate\,\vkb}, \D{\vc} = \vison \cdot \grayout{\vrate\,\vkc},\\
&\ \quad\qquad\D{\vrt} = \vison \cdot \grayout{\vrate\,\vkt}, \D{t}=1  \wedge t \leq \veps \land \va \geq 0 \land \vb \geq 0 \land \vc \geq 0\}\\[0.1in]
&(P \limply \dbox{\alpha}{Q}) \equiv (\grayout{\vconst  \land} \vrt \leq \vtmax \limply \dbox{\prepeat{\{\vctrl;\vode\}}}{\vrt \leq \vtmax})
\end{align*}
\caption{Bang-bang irreversible model safe}
\label{fig:irreversible-bangbang}
\end{figure}\vskip-0.1in
The greatest change is the addition of a \emph{time-triggered} controller:
the system now repeats in a loop, with the controller guaranteed to run at least every $\veps > 0$ time units.
The controller (\vctrl) is \emph{model-predictive} because it \emph{predicts} whether it would be dangerous to keep the reaction running for $\veps$ time.
If so, the reaction shuts off ($\passign{\vison}{0}$), else it turns on ($\passign{\vison}{1}$).
Note \vison is an \emph{indicator variable}; its only possible values are 0 and 1.
Specifically, the controller linearly predicts the maximum temperature change as $\veps\,\vrate\,\vkr$ and shuts off if the safe temperature would be exceeded. Importantly, this approach predicts unsafe events before they occur and shuts off before the damage is done.
Either way, the timer $t$ is reset to 0.

The $\vode$ is updated so that each reaction equation is multiplied by $\vison,$ causing no physical changes to occur when the reactor is turned off.
This model is best-suited for situations where it is possible to quickly halt a reaction. 
The $\vode$ gains an \emph{evolution domain constraint}, which serves to restrict its duration of evolution: an ODE may evolve only while the constraint remains true.
Our constraint serves two purposes.
Firstly, $t \leq \veps$ implements time-triggering: if each iteration takes at most $\veps$ time, there is at most $\veps$ delay between control cycles.
Secondly, the constraints $\va\geq 0 \land \vb\geq 0 \land \vc\geq 0$ model the physical assumption that concentrations cannot be negative.
For example, the reaction would end if $\va$ or $\vb$ reach zero.

Finally, the updated theorem statement $(P \limply \dbox{\alpha}{Q})$ is now a safety statement, stating that the reactor never exceeds its maximum safe temperature.

\paragraph{Proof.}
As the model now contains a loop, the proof uses \emph{loop invariant} reasoning in addition to \emph{differential invariant reasoning}, both distinct concepts from \emph{evolution domain constraints}.
We prove that the safety condition  $\vrt \leq \vtmax$ is a \emph{loop invariant}, meaning it holds before and after every loop repetition. We use the standard \textsf{loop} rule from \rref{sec:background-ddl}.

Already, a lemma arises in the ODE proof. 
Certain \emph{differential invariant} proofs can only succeed by first proving lemmas, called \emph{differential cut} formulas. which are then available as assumptions in the invariant proof.
Specifically, we prove the following cut:
\begin{itemize}
\item $\vtmax-\vrt >(\veps-t)\ \vrate\, \vkt$, meaning the remaining safe temperature gap exceeds the projected temperature change during the remaining time.
\end{itemize}
The cut proves automatically by differential invariant, from which the loop invariant, then safety condition, follow by automatic proof.

\subsubsection{Fixed Exponents}
For the next model, the first fundamental change is that we update the definition of $\vrate$ to use the current temperature, so that the reaction rate evolves exponentially over time.
Because dynamic reaction rates are an increase in complexity, we simply other aspects of the reaction rate formula by dropping $\vkra$ and $\vkrb$.
The remaining changes follow from that one: $\vamts$ is a helper definition to definitions such as $\vtaylorHi(x,t),$ which is an upper bound on temperature over time, constructed as a Taylor series approximation.
This use of a Taylor series approximation represents a fundamental change in proof approach for a fundamentally more complicated dynamics: for exponential dynamics, polynomial approximations are a crucial tool to simplify reasoning.
However, this Taylor bound is only provably an upper bound on a limited time interval which happens to be $1/(2\,\vamts),$ which we thus take as our upper limit on $\veps$.
In practice, we hypothesize that the time limit is artificial: time could be expressed in any desired units, increasing the interval.
The constants are updated to include assumptions on initial values of amounts and the controller is updated to use the Taylor approximation.
The $\vode$ is updated to explicitly assume nonnegative temperature, which is a safe assumption since our goal is to avoid high, not low, temperatures.
This new result shows safety with idealized modeling of concentrations under more realistic \emph{heating} assumptions.
\begin{figure}[!hbtp]
\begin{align*}
&\vrate \equiv \vrt\,\vainit\,\vbinit\ \ \veps \equiv 1/(2\,\vamts)\ \ \vamts \equiv \vkt\,\vainit\,\vbinit\ \ \vtaylorHi(x,t) \equiv (1 + 2\,t\,\vamts)\,x\\
&\vconst \equiv \vka > 0 \land \vkb > 0 \land \vkc > 0 \land \vkt > 0 \land \veps > 0 \land \vainit \geq 0 \land \vbinit \geq 0\\
&\vctrl \equiv \{\kwif(\vtmax \leq \vtaylorHi(\vrt,\veps))\grayout{\{\passign{\vison}{0}\}\kwelse\{\passign{\vison}{1}\}}\};\grayout{\passign{t}{0}}\\
&\vode \equiv \{\grayout{\D{\va} = \vison \cdot -\vrate\,\vka, \D{\vb} = \vison \cdot-\vrate\,\vkb, \D{\vc} = \vison \cdot \vrate\,\vkc,}\\
&\grayout{\ \quad\qquad\D{\vrt} = \vison \cdot \vrate\,\vkt, \D{t}=1 \wedge t \leq \veps \land \va \geq 0 \land \vb \geq 0 \land \vc \geq 0} \land \vrt \geq 0
\}\\[0.1in]
&(P \limply \dbox{\alpha}{Q}) \equiv (\vconst  \land \vrt > 0 \land \vrt \leq \vtmax \land \va = \vainit \land \vb = \vbinit \limply \dbox{\prepeat{\{\vctrl;\vode\}}}{\vrt \leq \vtmax})
\end{align*}
\caption{Bang-bang irreversible model safe with fixed exponent}
\label{fig:irreversible-fixedexp}
\end{figure}\vskip-0.1in

\paragraph{Proof.}
The loop invariant is unchanged.
We add several differential cuts; order matters since each one can serve as an assumption in following proofs:
\begin{itemize}
\item $t \geq 0$  just means time moves forward,
\item  $\vainit\,\vbinit\,\vrt\,\vkt \geq 0$ ensures forward reaction rate, and 
\item $\vtaylorHi(\vrt_{\textit{old}},t)-\vrt \geq 0$ bounds temperature $\vrt$ above with $\vtaylorHi()$ in terms of old temperature $\vrt_{\textit{old}}$.
\end{itemize}
The final cut requires advanced proof techniques because term  $\vtaylorHi(\vrt_{\textit{old}},t)-\vrt $ decreases; differential invariants alone are provably~\cite[Thm\ 6.1]{DBLP:journals/lmcs/Platzer12} insufficient for such terms.
The earliest suitable techniques required defining new (\emph{ghost}, or \emph{auxiliary}) variables in each proof, but constructing suitable definitions can be non-obvious in practice.
Fortunately, \KeYmaeraX has provided proof rules based on Darboux polynomial (inequality) reasoning~\cite[Corr.\ 3.2]{DBLP:journals/jacm/PlatzerT20} which can prove the same problems, but are higher-level:
\[\inferrule*[Left=$\text{dbx}_{\succcurlyeq}$]
{Q \limply \der{p} \geq g\,p}
{p \ensuremath{\succcurlyeq} 0 \limply \dbox{\pevolvein{\D{x}=f(x)}{Q}}{p\ensuremath{\succcurlyeq} 0}}\]
Here, both instances of $\succcurlyeq$ are replaced uniformly with one of $>$ or $\geq,$ where 
$\der{e}$ is the differential of $e,$ for polynomials $p, g$ where $p$ is called a \emph{Darboux} polynomial if the premise holds and $g$ is called its \emph{cofactor}.
It is natural to ask what power is gained by the addition of this proof rule.
Certainly it is stronger than differential invariant reasoning which would require $Q \limply \der{p} \geq 0$ because $g\,p$ are allowed to be negative.
Yet its full usefulness goes deeper, as the rule serves as a basis for differential radical invariant reasoning which is provably complete for semianalytic invariants~\cite[Thm.\ 4.5]{DBLP:journals/jacm/PlatzerT20}, a large class of invariants.

Darboux-based rules are complete for large classes of theorems, yet it is challenging to automatically find suitable polynomials in every case. 
For our example model, \KeYmaeraX did not find a suitable polynomial, but performing algebra by hand did result in a suitable polynomial: using the definition of the ODE, solve for a polynomial that satisfies the proof goal, in this case: $g \equiv \vainit\,\vbinit\,\vkt$.
After choosing a suitable Darboux polynomial, the remaining proof goals completed using \KeYmaeraX's default proof method.
Further applications of Taylor approximations are discussed in \rref{sec:related-work}.

\subsubsection{Dynamic Exponents}
Even our final controlled model, below, makes some important simplifying assumptions. 
Note that our model makes the impact of temperature on reaction rate a linear one, whereas the true 
Arrhenius equation~\cite[Eq.\ 5.1]{hellgardt2000engineering} implies an exponential effect on reaction rate.
Linear functions can locally approximate exponential ones, but exponentials remain of future interest.
Despite these limitations, the final model is important because it shows safety with both non-trivial heating dynamics \emph{and} nontrivial concentration dynamics.

The core change in the final model is a more advanced reaction rate dynamics, where the reaction rate dynamically changes in response to the concentration of each reactant.
Definitions $\vamts$ and $\veps$ are updated for the same reason.
The timestep $\veps$ now changes dynamically: as the reaction proceeds, the acceptable delay \emph{increases}, thus becoming easier to satisfy.
It simplifies the analysis to have $\veps$ change only at each loop iteration rather than continuously, so we introduce variables $\vamid, \vbmid$ to stand for the values of $\va,\vb$ at the \emph{start} of each ODE evolution.
The changes to the model are modest, but the dynamic changes are notable: the reaction rate is now a product of three changing variables, no longer an exponential with a fixed base.
Likewise, additional proof steps will be required to account for changing concentrations, but the core proof approach is unchanged.%
\begin{figure}[!hbtp]\vskip-0.1in
\begin{align*}%
&\vrate \equiv \vrt\,\va\,\vb\ \ \veps \equiv 1/(2\,\vamid\,\vbmid\,\vkt)\ \ \vamts \equiv \va\,\vb\,\vkt\ \ \vctrl\\
&\grayout{\vconst \equiv \vka > 0 \land \vkb > 0 \land \vkc > 0 \land \vkt > 0 \land \veps > 0 \land \vainit \geq 0 \land \vbinit \geq 0}\\
&\vctrl \equiv \grayout{\{\kwif(\vtmax \leq \vtaylorHi(\vrt,\veps))\{\passign{\vison}{0}\}\kwelse\{\passign{\vison}{1}\}\};\passign{t}{0}}; \passign{A_1}{A}; \passign{B_1}{B}\\
&\grayout{\vode \equiv \{\D{\va} = \vison \cdot -\vrate\,\vka, \D{\vb} = \vison \cdot-\vrate\,\vkb, \D{\vc} = \vison \cdot \vrate\,\vkc,}\\
&\grayout{\ \quad\qquad\D{\vrt} = \vison \cdot \vrate\,\vkt, \D{t}=1  \wedge t \leq \veps \land \va \geq 0 \land \vb \geq 0 \land \vc \geq 0 \land \vrt \geq 0\}}\\
&\grayout{(P \limply \dbox{\alpha}{Q}) \equiv (\vconst  \land \vrt > 0 \land \vrt \leq \vtmax \land \va = \vainit \land \vb = \vbinit \limply \dbox{\prepeat{\{\vctrl;\vode\}}}{\vrt \leq \vtmax})}
\end{align*}%
\caption{Bang-bang irreversible model safe with dynamic exponent}%
\label{fig:irreversible-dynexp}%
\end{figure}%

\paragraph{Proof.}
In this proof, the reaction rate changes as the concentration of each reactant changes, so we strengthen the loop invariant
to capture the status of the reactant concentrations:
$0\leq \vrt\land\vrt\leq\vtmax\land \va \leq\vainit \land \vb \leq\vbinit$.
The differential cuts are similar to before, with an additional lemma that the concentrations of the first two reactants decrease:
$\va\leq\vamid \land \vamid \leq \vainit \land \vb\leq\vbmid \land \vbmid \leq \vbinit$.
The differential cut for the Taylor series is unchanged, and the same Darboux polynomial 
$g \equiv \vainit\,\vbinit\,\vkt$ suffices.

\subsection{Uncontrolled Reversible Reactions}
\label{sec:reversible}
We study reversible reactions, which are crucial to society.
For example, ammonia synthesis is critical to modern agriculture~\cite{liu2014ammonia}.
We consider a textbook scenario where two reactants $A$ and $B$ can each react to form the other:
\[A \rightleftharpoons B\]
To our knowledge, we provide the first computer-checked proofs for the asymptotic behavior of this classic, widely-used textbook scenario.
Specifically, our final model shows \emph{persistence}~\cite{sogokon2017verifying}, a relative of stability: the system eventually gets arbitrarily close to its equilibrium state, then stays close forever.
We build up to this result with lemmas: the system is always moving toward equilibrium and can arbitrarily approach equilibrium in finite, bounded time.
To complete the story, we show that although the equilibrium can always be arbitrarily approximated, it can never be reached exactly.


\subsubsection{Pure Reactant Decreases}
We consider a scenario where we start with pure reactant $A,$ which then becomes a mixture.
We show the current amount of $A$ never exceeds the initial amount, which is intuitive by conservation of mass.
The lemma might be of practical use in its own right, e.g., to verify that a container never overflows, but we mainly use the lemma as a   building block for persistence.
\begin{figure}[!hbtp]\vskip-0.1in%
\begin{align*}%
&\vode \equiv \pevolve{\D{\va} = -\va\,\vkf + \vb\,\vkr, \D{\vb}=\va\,\vkf  -\vb\,\vkr}\\
&\vconst \equiv  \vainit > 0  \land \vkr > 0 \land \vkf > 0\\
&(P \limply \dbox{\alpha}{Q}) \equiv (\vconst  \land \va = \vainit\land \vb = 0 \limply \dbox{\vode}{\va \leq \vainit})%
\end{align*}%
\caption{Reversible model decreases $\va$}%
\label{fig:reversible-basic}%
\end{figure}%
Here, the two reactants are named $\va$ and $\vb$, with initial values $\va = \vainit > 0$ and $\vb = 0$.
Reactants $\va$ and $\vb$ are engaged in a \emph{reversible reaction} where $\va$ converts to $\vb$ at forward rate $\vkf$ and $\vb$ converts to $\va$ at reverse rate $\vkr$.
It is well-known~\cite[Ch.\ 3]{hellgardt2000engineering} that the system asymptotically approaches an equilibrium state, called a \emph{dynamic equilibrium}, in which the forward and reverse reactions perfectly cancel out.
We define $\vode$ using a classic textbook model of a reversible reaction, which does not model heat: the reaction rates are based solely on concentrations and constants.

\paragraph{Proof.}
This proof completes automatically: the automatic prover successfully reasons by differential invariant.

\subsubsection{Equilibrium Avoidance}
We show that the amounts of the reactants never exactly reach the equilibrium.
Though not directly used in the persistence proof, we prove this because it is a fundamental property in its own right which tacitly influences how a chemical plant is designed and operated.
An operator would never wait for perfect equilibrium to occur, only for the system to get \emph{close} to equilibrium, because perfect equilibrium (provably) never occurs.

The initial condition and ODE are unchanged, only the postcondition changes, which mandates a new proof approach.
To state the new postcondition, we define the amounts  $\vequiA$ and $\vequiB$ of $\va$ and $\vb$ present at the equilibrium.
\begin{figure}[!hbtp]
\begin{align*}%
&\grayout{\vode \equiv \pevolve{\D{\va} = -\va\,\vkf + \vb\,\vkr, \D{\vb}=\va\,\vkf  -\vb\,\vkr}}\\
&\grayout{\vconst \equiv  \vainit > 0  \land \vkr > 0 \land \vkf > 0}\\
&\vequiA \equiv \vainit\,(\vkr/(\vkf + \vkr)) \quad\vequiB \equiv \vainit\,(\vkf/(\vkf + \vkr))\\
&(P \limply \dbox{\alpha}{Q}) \equiv (\grayout{\vconst  \land \va = \vainit\land \vb = 0 \limply }\dbox{\grayout{\vode}}{\va \neq \vequiA})%
\end{align*}
\caption{Reversible model never at equilibrium}%
\label{fig:reversible-avoid}%
\end{figure}
The above definitions of $\vequiA$ and $\vequiB$ can be found by solving for 
equilibrium ($\D{\va}=0 \land \D{\vb}=0$)  in $\vode$ subject to conservation of mass ($\va + \vb = \vainit$).

\paragraph{Proof.}
A simple change in postcondition creates a major increase in proof complexity, because we now wish to show a lower bound instead of an upper bound.
We use multiple differential cuts, one of which uses Darboux reasoning.
\begin{itemize}
\item $\va-\vainit\,(\vkr/(\vkf+\vkr))>0$ means $\va$'s rate of change is always in the direction of the equilibrium
\item $\va+\vb=\vainit$ is conservation of mass
\item $\va>0\land \vb\geq 0$ means we never have a negative amount of either reactant, the first being positive.
This requires a Darboux argument with polynomial $-(\vkf+\vkr)$ because the amount of the first reactant does decrease with time.
\end{itemize}
Once these cuts are proved, automation suffices to finish the proof.

\subsubsection{Equilibrium Approach}
We show that we get arbitrarily close to the equilibrium, given sufficient time. 
For every positive epsilon ($\veps > 0$), there exists a time when we get that close to the equilibrium.
The assumption changes slightly; the theorem statement changes more: we prove a \emph{diamond} modality $\ddiamond{\vode}{\va \leq \vequiA + \veps}$ because we want to show we \emph{eventually} approach the equilibrium.
The practical impact of this result is that if an engineer desires an almost-perfect equilibrium, that can be attained, but the cost is time.
\begin{figure}[!hbtp]
\begin{align*}
&\vconst \equiv  \grayout{\vainit > 0  \land \vkr > 0 \land \vkf > 0 \land} \veps > 0\\
&\grayout{\vode \equiv \pevolve{\D{\va} = -\va\,\vkf + \vb\,\vkr, \D{\vb}=\va\,\vkf  -\vb\,\vkr}}\\
&\grayout{\vequiA \equiv \vainit\,(\vkr/(\vkf + \vkr)) \quad\vequiB \equiv \vainit\,(\vkf/(\vkf + \vkr))}\\
&(P \limply \ddiamond{\alpha}{Q}) \equiv (\grayout{\vconst  \land \va = \vainit\land \vb = 0 \limply} \ddiamond{\vode}{\va \leq \vequiA + \veps})
\end{align*}\vskip-0.1in
\caption{Reversible model approaches equilibrium}\vskip-0.1in
\label{fig:reversible-approach}
\end{figure}

\paragraph{Proof.}
 Previous proofs highlighted advances in proof automation for box properties of ODEs;
 this proof relies on advances in proof automation for diamond properties of ODEs.
 A \emph{differential variant} proof is the diamond counterpart to \emph{differential invariant} reasoning for box properties.
 The \emph{differential variant} principle~\cite[Corr.\ 24]{DBLP:journals/fac/TanP21} says: if there is a lower bound on the rate of progress we make toward our goal at all times, we will get there eventually.
 \[
 \inferrule*[Left={$\text{dV}\stackrel{\exists}{\succcurlyeq}$}]
 {\exists d > 0 \forall x (\lnot (p \geq 0) \limply \der{p} \geq d)}
 {\ddiamond{\pevolve{\D{x}=f(x)}}{p \succcurlyeq 0}}
 \]
 where $\succcurlyeq$ stands for either $>$ or $\geq,$ where $d$ is a fresh variable and where $\D{x}=f(x)$ provably has a global solution (i.e., for all time).

The key insight behind our proof is that the rate of progress is proportional to our current displacement from the equilibrium.
Since we seek to get the displacement within some $\veps$, we can assume without loss of generality that the current displacement is at least $\veps$, giving a bound $d$ on the progress rate: 
$d = \veps\,(\vkf + \vkr)$.
This progress rate also confirms standard intuitions about the system dynamics: higher rates of progress are made when far away from the equilibrium and when reaction rates are high.

\subsubsection{Persistence}
\label{sec:reverse-stability}
Persistence means there exists a point after which we forever remain within \verb|eps| of the equilibrium.
Persistence is of practical importance because it shows both the system can get arbitrarily close to equilibrium \emph{and} that the system stays that way \emph{indefinitely}.
In short, this result is important from a control perspective because it shows the system is well-controlled, even without a controller
As a theorem-proving case study, persistence is an excellent comprehensive test case because it combines boxes and diamonds.
Only the theorem statement need be updated; all other definitions are unchanged:
\begin{figure}[!hbtp]
\centering
\begin{align*}
&\grayout{\vconst \equiv  \vainit > 0  \land \vkr > 0 \land \vkf > 0 \land} \veps > 0\\
&\grayout{\vode \equiv \pevolve{\D{\va} = -\va\,\vkf + \vb\,\vkr, \D{\vb}=\va\,\vkf  -\vb\,\vkr}}\\
&\grayout{\vequiA \equiv \vainit\,(\vkr/(\vkf + \vkr)) \quad\vequiB \equiv \vainit\,(\vkf/(\vkf + \vkr))}\\
&(P \limply \ddiamond{\alpha}{Q}) \equiv (\grayout{\vconst  \land \va = \vainit\land \vb = 0 \limply} \ddiamond{\vode}{\dbox{\vode}{\va \leq \vequiA + \veps}})
\end{align*}
\caption{Reversible model is stable.}
\label{fig:reversible-stable}
\end{figure}%
\vskip-0.2in\paragraph{Proof.}
We combine proof techniques, first showing we eventually approach the equilibrium (variant reasoning), then showing the concentration of $A$ never increases again (invariant reasoning).

A major strength of logic is \emph{compositionality}: complex proofs are but combinations of simple parts. 
A \dL proof of $\ddiamond{\alpha}{\dbox{\alpha}{P}}$ can be divided into a variant proof and invariant proof, for example. 
At a high level, \KeYmaeraX lived up to this compositionality promise.
At a low level, there is always room for improvement: the $\dbox{\alpha}{P}$ proof assumes $\vconst,$ i.e., it assumes constants never change.
Due to limitations of the differential variant rule, we had to prove the constants never change, albeit with a simple proof.
The limitation appears incidental to \KeYmaeraX's implementation, not fundamental.
It speaks well of the implementation used in these case studies that this was the only instance where the automation added new proof challenges.
This serves as a reminder that theorem-proving case studies are dually important, showing both the gains from new automation and which features deserve future optimization.

\section{Related Work}
\label{sec:related-work}
Related work includes hybrid systems verification, reactor design, and reaction kinetics.
We begin with theorem-proving approaches to verification, specifically.
\paragraph{Hybrid Systems Theorem Proving.}
Specialized \emph{hybrid systems} theorem-provers~\cite{DBLP:conf/cade/FultonMQVP15,DBLP:conf/icfem/WangZZ15} provide a high degree of generality and rigor, while making efforts to mitigate the high degree of user effort typical of theorem-proving.
For example, generality in our case study means many different reactions and reactors are supported by modifying parameter values, with no new proof effort.
Rigor is not merely of theoretical interest:
in many hybrid systems reasoning techniques which do not share our rigorous logical foundations,
many soundness edge cases have recently been identified~\cite[Tab.\ 1]{DBLP:journals/fac/TanP21}.
Soundness violations are unacceptable in verification.

We use the \KeYmaeraX~\cite{DBLP:conf/cade/FultonMQVP15} prover for its exceptional rigor: its axioms have been proved sound in a theorem-prover~\cite{DBLP:conf/cpp/BohrerRVVP17}
and it soundly derives its advanced proof methods~\cite[Tab.\ 1]{DBLP:journals/fac/TanP21,DBLP:journals/jacm/PlatzerT20} from sound axioms.

Hybrid Hoare Logic (HHL)~\cite{DBLP:conf/aplas/LiuLQZZZZ10,DBLP:conf/icfem/WangZZ15} is another notable hybrid prover; an HHL case study similar to ours could be interesting future work.
HHL Prover and \KeYmaeraX both base their ODE invariant automation on the same core algorithm~\cite{DBLP:conf/emsoft/LiuZZ11}, so this aspect of automation is likely comparable in both.

\paragraph{Other Logical Approaches}
We are aware of only one prior logical proof~\cite{DBLP:conf/hybrid/QiwenW95} of a chemical process with nontrivial hybrid dynamics. 
Unlike ours, it is not in a theorem-prover and does not address persistence nor reactions, but rather a mixing process.
General-purpose theorem-provers~\cite{DBLP:conf/iceccs/Abraham-MummSH01,DBLP:journals/tecs/DupontASP21,DBLP:conf/tphol/MitraC08,DBLP:conf/cpp/Rouhling18} have formalized hybrid systems, including stability~\cite{DBLP:conf/cpp/Rouhling18,DBLP:conf/tphol/MitraC08}, but not applied them to reactions.

\paragraph{Reachability}
Model-checkers based on reachability analysis~\cite{DBLP:conf/cav/ChenAS13,DBLP:conf/adhs/AlthoffGK18,DBLP:conf/hybrid/DuggiralaPM015,DBLP:conf/cav/FrehseGDCRLRGDM11} are the primary competitors to hybrid systems theorem-provers.
They provide greater automation at the cost of accepting restrictions in generality.
Details vary, but common restrictions include special-case guarantees (is a \emph{specific} reaction safe?), time-bounded analyses (am I safe \emph{for a time}?) or conservative approximations of dynamics. Their trusting computing base is typically larger than a theorem-prover's, complicating rigor.

Taylor approximations, particularly Taylor models~\cite{DBLP:journals/rc/BerzH98}, are broadly useful in reachability analysis, e.g., in Flow*~\cite{DBLP:conf/cav/ChenAS13} and CORA~\cite{DBLP:conf/adhs/AlthoffGK18}.
We have shown that Taylor approximations are equally useful in \KeYmaeraX, where they come with proofs.

\paragraph{Stability and Persistence}
Hybrid system stability is well-studied both inside~\cite{DBLP:conf/tacas/TanP21,DBLP:conf/tphol/MitraC08,DBLP:conf/cpp/Rouhling18} and outside~\cite{DBLP:journals/deds/KoutsoukosHLA98,DBLP:conf/cdc/MitraL04,lozano2000stabilization} theorem-provers, with persistence also studied~\cite{sogokon2017verifying}.
Lyapunov functions have shown stability of a chemical reaction on paper, but not in a prover~\cite{DBLP:journals/entropy/Hangos10}.
Stability and its relatives in \KeYmaeraX specifically are a new topic~\cite{DBLP:conf/tacas/TanP21}; we contribute the first worked \KeYmaeraX case study for an application of industrial interest.

\paragraph{Chemical Engineering.}
The chemical engineering results we formalized are classical; our innovation is the generality and rigor with which we formalize them in \KeYmaeraX.
Standard textbooks provided kinetics for well-mixed adiabatic batch reactors~\cite[Eq.\ 2.93]{hellgardt2000engineering}, uncontrolled reversible reactions~\cite[Ch.\ 3]{hellgardt2000engineering},
and the Arrhenius equation~\cite[Eq.\ 5.1]{hellgardt2000engineering}.
Standard control theory textbooks introduce model-predictive control and bang-bang control~\cite{glad2018control}.

Although basic models of reactors are widely-used in formal methods, ours is the first in a  theorem-prover. 
It additional overcomes others' limitations:
\begin{itemize}
\item Previous chemical proofs ignored persistence and reactors~\cite{DBLP:conf/hybrid/QiwenW95}
\item Optimal scheduling~\cite{ctx4405807090004746} and safety arguments~\cite{lukoschus2004compositional} have used simplistic finite state machines
\item A verified plant design used simple piecewise-constant dynamics~\cite{hassapis1998validation}
\item CEGAR verification of tanks~\cite{DBLP:series/asc/NellenAW15} ignored reactors
\end{itemize}

Though we build on such broad related work, our contribution of general-purpose proofs about chemical reactors and reactions in a theorem-prover fills a significant gap in the verification literature.


\section{Conclusion}
\label{sec:conclusion}
We used the \KeYmaeraX theorem prover for differential dynamic logic to formalize two case studies: a batch reactor and a reversible reaction, each of which consisted of four models and their proofs.
This work served two purposes:
\begin{itemize}
\item To our knowledge, we provide the first proof in a theorem prover of these foundational chemical engineering results
\item We demonstrate how recent advances in \KeYmaeraX's automation, such as its implementation invariant checking, Darboux reasoning, and differential variants, contribute to the proofs
\end{itemize}
One direction for future work is verifying reactors with more advanced controllers such as PID (proportional-integral-derivative) controllers~\cite[Ch.\ 13]{cpc}.
However, potential future work is broad in nature, reaching well beyond chemical reactor design.
Techniques such as invariant checking and Taylor series are of general applicability using various tools, though
\KeYmaeraX provides a rigorous implementation of both.
Differential variants are widely useful for proving ODE properties that are true eventually, but not at every moment.
We have shown one significant application for all these proof techniques; their are certainly others because the applications of hybrid systems models are diverse.

\bibliographystyle{splncs04}
\bibliography{chemsyn}

\begin{thebibliography}{10}
\providecommand{\url}[1]{\texttt{#1}}
\providecommand{\urlprefix}{URL }
\providecommand{\doi}[1]{https://doi.org/#1}

\bibitem{DBLP:conf/iceccs/Abraham-MummSH01}
{\'{A}}brah{\'{a}}m{-}Mumm, E., Steffen, M., Hannemann, U.: Verification of
  hybrid systems: Formalization and proof rules in {PVS}. In: {ICECCS}. {IEEE}
  (2001). \doi{10.1109/ICECCS.2001.930163}

\bibitem{DBLP:conf/adhs/AlthoffGK18}
Althoff, M., Grebenyuk, D., Kochdumper, N.: Implementation of taylor models in
  {CORA} 2018. In: {ARCH}. {EPiC Series in Computing}, vol.~54. EasyChair
  (2018). \doi{10.29007/zzc7}

\bibitem{bauer2000case}
Bauer, N., Kowalewski, S., Sand, G., L{\"o}hl, T.: A case study: Multi product
  batch plant for the demonstration of control and scheduling problems. In:
  {ADPM} (2000)

\bibitem{DBLP:journals/rc/BerzH98}
Berz, M., Hoffst{\"{a}}tter, G.: Computation and application of {Taylor}
  polynomials with interval remainder bounds. Reliab. Comput.  \textbf{4}(1)
  (1998). \doi{10.1023/A:1009958918582}

\bibitem{DBLP:conf/cpp/BohrerRVVP17}
Bohrer, R., Rahli, V., Vukotic, I., V{\"o}lp, M., Platzer, A.: Formally
  verified differential dynamic logic. In: {CPP}. {ACM} (2017).
  \doi{10.1145/3018610.3018616}

\bibitem{DBLP:conf/cav/ChenAS13}
Chen, X., {\'{A}}brah{\'{a}}m, E., Sankaranarayanan, S.: Flow*: An analyzer for
  non-linear hybrid systems. In: Sharygina, N., Veith, H. (eds.) {CAV}. {LNCS},
  vol.~8044. Springer (2013). \doi{10.1007/978-3-642-39799-8_18}

\bibitem{DBLP:conf/hybrid/DuggiralaPM015}
Duggirala, P.S., Potok, M., Mitra, S., Viswanathan, M.: {C2E2:} a tool for
  verifying annotated hybrid systems. In: {HSCC}. {ACM} (2015).
  \doi{10.1145/2728606.2728646}

\bibitem{DBLP:journals/tecs/DupontASP21}
Dupont, G., Ameur, Y.A., Singh, N.K., Pantel, M.: Event-b hybridation: {A}
  proof and refinement-based framework for modelling hybrid systems. {ACM}
  Trans. Embed. Comput. Syst.  \textbf{20}(4) (2021). \doi{10.1145/3448270}

\bibitem{DBLP:conf/cav/FrehseGDCRLRGDM11}
Frehse, G., Guernic, C.L., Donz{\'{e}}, A., Cotton, S., Ray, R., Lebeltel, O.,
  Ripado, R., Girard, A., Dang, T., Maler, O.: Spaceex: Scalable verification
  of hybrid systems. In: {CAV}. {LNCS}, vol.~6806. Springer (2011).
  \doi{10.1007/978-3-642-22110-1_30}

\bibitem{DBLP:conf/itp/FultonMBP17}
Fulton, N., Mitsch, S., Bohrer, R., Platzer, A.: Bellerophon: Tactical theorem
  proving for hybrid systems. In: Ayala{-}Rinc{\'{o}}n, M., Mu{\~{n}}oz, C.A.
  (eds.) {ITP}. {LNCS}, vol. 10499. Springer (2017).
  \doi{10.1007/978-3-319-66107-0_14}

\bibitem{DBLP:conf/cade/FultonMQVP15}
Fulton, N., Mitsch, S., Quesel, J., V{\"{o}}lp, M., Platzer, A.: Keymaera {X:}
  an axiomatic tactical theorem prover for hybrid systems. In: {CADE}. {LNCS},
  vol.~9195. Springer (2015). \doi{10.1007/978-3-319-21401-6_36}

\bibitem{glad2018control}
Glad, T., Ljung, L.: Control theory. {CRC} Press (2018)

\bibitem{DBLP:journals/entropy/Hangos10}
Hangos, K.M.: Engineering model reduction and entropy-based lyapunov functions
  in chemical reaction kinetics. Entropy  \textbf{12}(4) (2010).
  \doi{10.3390/e12040772}

\bibitem{hassapis1998validation}
Hassapis, G., Kotini, I., Doulgeri, Z.: Validation of a {SFC} software
  specification by using hybrid automata. {IFAC} Proc.  \textbf{31}(15) (1998)

\bibitem{DBLP:journals/deds/KoutsoukosHLA98}
Koutsoukos, X.D., He, K.X., Lemmon, M.D., Antsaklis, P.J.: Timed petri nets in
  hybrid systems: Stability and supervisory control. Discret. Event Dyn. Syst.
  \textbf{8}(2) (1998). \doi{10.1023/A:1008293802713}

\bibitem{liu2014ammonia}
Liu, H.: Ammonia synthesis catalyst 100 years: Practice, enlightenment and
  challenge. Chinese Journal of Catalysis  \textbf{35}(10),  1619--1640 (2014)

\bibitem{DBLP:conf/aplas/LiuLQZZZZ10}
Liu, J., Lv, J., Quan, Z., Zhan, N., Zhao, H., Zhou, C., Zou, L.: A calculus
  for hybrid {CSP}. In: {APLAS}. {LNCS}, vol.~6461. Springer (2010).
  \doi{10.1007/978-3-642-17164-2_1}

\bibitem{DBLP:conf/emsoft/LiuZZ11}
Liu, J., Zhan, N., Zhao, H.: Computing semi-algebraic invariants for polynomial
  dynamical systems. In: Chakraborty, S., Jerraya, A., Baruah, S.K.,
  Fischmeister, S. (eds.) {EMSOFT}. {ACM} (2011). \doi{10.1145/2038642.2038659}

\bibitem{lozano2000stabilization}
Lozano, R., Fantoni, I., Block, D.J.: Stabilization of the inverted pendulum
  around its homoclinic orbit. Systems \& Control Letters  \textbf{40}(3)
  (2000)

\bibitem{lukoschus2004compositional}
Lukoschus, B.: Compositional verification of industrial control systems:
  methods and case studies. Ph.D. thesis, Christian-Albrechts Universit{\"a}t
  Kiel (2004)

\bibitem{DBLP:conf/tphol/MitraC08}
Mitra, S., Chandy, K.M.: A formalized theory for verifying stability and
  convergence of automata in {PVS}. In: {TPHOLs}. {LNCS}, vol.~5170. Springer
  (2008). \doi{10.1007/978-3-540-71067-7_20}

\bibitem{DBLP:conf/cdc/MitraL04}
Mitra, S., Liberzon, D.: Stability of hybrid automata with average dwell time:
  an invariant approach. In: {CDC}. {IEEE} (2004).
  \doi{10.1109/CDC.2004.1430238}

\bibitem{DBLP:conf/fide/MitschP16}
Mitsch, S., Platzer, A.: The {KeYmaera X} proof {IDE}: Concepts on usability in
  hybrid systems theorem proving. In: Dubois, C., Masci, P., M{\'{e}}ry, D.
  (eds.) {FIDE}. {EPTCS}, vol.~240, pp. 67--81 (2016).
  \doi{10.4204/EPTCS.240.5}

\bibitem{DBLP:series/asc/NellenAW15}
Nellen, J., {\'{A}}brah{\'{a}}m, E., Wolters, B.: A {CEGAR} tool for the
  reachability analysis of plc-controlled plants using hybrid automata. In:
  Bouabana{-}Tebibel, T., Rubin, S.H. (eds.) {FMI}, {AISC}, vol.~346. Springer
  (2015). \doi{10.1007/978-3-319-16577-6_3}

\bibitem{DBLP:journals/lmcs/Platzer12}
Platzer, A.: The structure of differential invariants and differential cut
  elimination. Logical Methods in Computer Science  \textbf{8}(4) (2012).
  \doi{10.2168/LMCS-8(4:16)2012}

\bibitem{Platzer18}
Platzer, A.: Logical Foundations of Cyber-Physical Systems. Springer, Cham
  (2018). \doi{10.1007/978-3-319-63588-0}

\bibitem{DBLP:journals/jacm/PlatzerT20}
Platzer, A., Tan, Y.K.: Differential equation invariance axiomatization. J.
  {ACM}  \textbf{67}(1) (2020). \doi{10.1145/3380825}

\bibitem{ctx4405807090004746}
Potočnik, B., Bemporad, A., Torrisi, F.D., Mušič, G., Zupančič, B.: Hybrid
  modelling and optimal control of a multiproduct batch plant. Control
  Engineering Practice  \textbf{12}(9) (2004)

\bibitem{DBLP:conf/cpp/Rouhling18}
Rouhling, D.: A formal proof in {Coq} of a control function for the inverted
  pendulum. In: {CPP}. {ACM} (2018). \doi{10.1145/3167101}

\bibitem{hellgardt2000engineering}
Schmidt, L.D.: The engineering of chemical reactions (1998)

\bibitem{sogokon2017verifying}
Sogokon, A., Jackson, P.B., Johnson, T.T.: Verifying safety and persistence
  properties of hybrid systems using flowpipes and continuous invariants. In:
  {NFM}. pp. 194--211. Springer (2017)

\bibitem{cpc}
Stephanopoulos, G.: Chemical Process Control: An Introduction to Theory and
  Practice. Prentice-Hall (1984)

\bibitem{DBLP:journals/fac/TanP21}
Tan, Y.K., Platzer, A.: An axiomatic approach to existence and liveness for
  differential equations. Formal Aspects Comput.  \textbf{33}(4) (2021).
  \doi{10.1007/s00165-020-00525-0}

\bibitem{DBLP:conf/tacas/TanP21}
Tan, Y.K., Platzer, A.: Deductive stability proofs for ordinary differential
  equations. In: Groote, J.F., Larsen, K.G. (eds.) {TACAS}. {LNCS}, vol. 12652,
  p. 181–199. Springer (2021). \doi{10.1007/978-3-030-72013-1_10}

\bibitem{DBLP:conf/icfem/WangZZ15}
Wang, S., Zhan, N., Zou, L.: An improved {HHL} prover: An interactive theorem
  prover for hybrid systems. In: {ICFEM}. {LNCS}, vol.~9407. Springer (2015).
  \doi{10.1007/978-3-319-25423-4_25}

\bibitem{DBLP:conf/hybrid/QiwenW95}
Xu, Q., He, W.: Hierarchical design of a chemical concentration control system.
  In: Hybrid Systems. {LNCS}, vol.~1066. Springer (1995).
  \doi{10.1007/BFb0020952}

\end{thebibliography}

\end{document}